# Scintillation Properties and Electronic Structure of Intrinsic and Extrinsic Mixed Elpasolites $Cs_2NaRBr_3I_3$ ($R$ = La, Y)

Hua Wei[1,*], Mao-Hua Du[2], Luis Stand[1], Zhao Zhao[3],
Hongliang Shi[2], Mariya Zhuravleva[1], Charles L. Melcher[1]

[1]*Scintillation Materials Research Center, Department of Materials Science and Engineering,
University of Tennessee, Knoxville, Tennessee 37996, USA*
[2]*Advanced Materials Group, Materials Science & Technology Division, Oak Ridge National Laboratory,
Oak Ridge, Tennessee 37831, USA*
[3]*Department of Physics, Stanford University, Stanford, California 94305, USA*

\* sundyhw@gmail.com, current address: Radiation Monitoring Devices Inc., 44 Hunt Street, Watertown, MA, 02472, USA

Scintillators attract wide research interest for their distinct applications in radiation detection. Elpasolite halides are among the most promising scintillators due to their high structural symmetry and good scintillation performance. Better understanding of their underlying scintillation mechanism opens up new possibilities in scintillator development. In this work, we employ a variety of experimental techniques to study the two mixed-anion elpasolites $Cs_2NaRBr_3I_3$ ($R$ = La, Y). The emission of intrinsic $Cs_2NaRBr_3I_3$ with a light yield ranging from 20,000 to 40,000 ph/MeV is dominant by self-trapped exciton emission. Partial substitution of $R$ with Ce introduces a competing emission, the $Ce^{3+}$ $5d$ to $4f$ radiative transition. *Ab initio* calculations were performed to investigate the electronic structures as well as the binding energies of polarons in $Cs_2NaRBr_6$. The calculated large self-trapped exciton binding energies are consistent with the observed high light yield due to self-trapped exciton emission. The unique electronic structure of halide elpasolites as calculated enhances the STE stability and the STE emission. The highly tunable scintillation properties of mixed-anion elpasolites underscore the role of their complex scintillation mechanism. Our study provides guidance for the design of new elpasolites scintillators with exceptional energy resolution and light yield desirable for applications.

## I. INTRODUCTION

Scintillators absorb and convert high-energy photons or particles into multiple low-energy photons [1]. They are widely used for x-ray, gamma ray, neutron, and charged particle detection. Positron Emission Tomography (PET), a state-of-art nuclear imaging tool to examine the body metabolism for early-stage cancer diagnosis [2], its ultimate performance is strongly tied to the properties of scintillators [3-5]. LSO ($Lu_2SiO_5$:Ce) or its analogue LYSO [6] are currently employed in PET. However, the imaging quality suffers most from their low light yield. Besides, the shortage of the raw material $Lu_2O_3$ and the substantial increase of cost urge the industry to seek better performance scintillators with less cost. In high-energy physics, scintillators with fast timing and high density are desirable for designing next generation hadron colliders [7]. In oil well logging, scintillators are used to measure the radioactivity of the clay formation, as well as conduct the elemental analysis.

The workhorse NaI:Tl is replaced by new $LaBr_3$:Ce, which yields more precise data with shorter response time [8].

Over the past few decades, halide compounds have shown great potential as the next generation scintillators [4,9-11]. For instance, $LaBr_3$:Ce achieves a state-of-art coincidence resolving time of 100 ps for time-of-flight PET [12]. $KSr_2I_5$:Eu possesses an exceptional energy resolution of 2.4% at 662 keV [13]. $Cs_2LiYCl_6$:Ce has efficient neutron/gamma ray pulse shape discrimination (PSD) ability for nuclear non-proliferation applications [14]. Among the numerous metal halides, elpasolite halides with highly symmetric crystal structures are of particular interest [15-18]. Their cubic or pseudo-cubic isotropic structures reduce the impact of thermo-mechanical stress during crystal growth, thus minimize the cracks and improve the production yields [19].

The discovery of elpasolite halides can be tracked back to 1883, where $K_2NaAlF_6$ was first identified in minerals [20]. Since then, numerous elpasolites have been reported. The halide elpasolites have a general form of $A_2BRX_6$, where A and B are monovalent alkali metals, $R$ is a trivalent rare earth element, and X is the halogen element. The elpasolite structure can be viewed as a cationic-ordered perovskite (CaTiO$_3$-type) structure [21]. In the ideal situation, the elpasolite structure has Fm$\bar{3}$m cubic symmetry.

According to Goldschmidt [22,23], the Goldschmidt tolerance factor of elpasolites can be expressed as:

$$t = \frac{R_A + R_X}{\sqrt{2}[\frac{1}{2}(R_B + R_R) + R_X]} \quad (1)$$

Where t represents the Goldschmidt tolerance factor, $R_A$, $R_B$, $R_R$, and $R_X$ are the ionic radii of A$^+$, B$^+$, $R^{3+}$, and X$^-$.

Based on Eq. (1), for the mixed-anion system in this work, the Goldschmidt tolerance factor can be written as:

$$t = \frac{R_A + \overline{R_X + R_{X'}}}{\sqrt{2}[\frac{1}{2}(R_B + R_R) + \overline{R_X + R_{X'}}]} \quad (2)$$

Where $\overline{R_X + R_{X'}}$ is the average ionic radius of the two halogen anions X$^-$ and X'$^-$. As the tolerance factor approaches unity, the crystal structure is more likely to be cubic [21,23,24].

Doty *et al.* [25] considered 640 potential halide elpasolites as scintillation host materials. Among the large number of elpasolites, only a few of them have been experimentally proved as scintillators. Cs$_2$LiYCl$_6$:Ce is one of the first elpasolite scintillators successfully applied in neutron/gamma ray detection [15,26]. Cs$_2$LiYCl$_6$:Ce has a cubic crystal structure with a band gap of 6.8-7.5 eV [27]. The light yield for gamma ray excitation is around 20,000 ph/MeV. Cs$_2$LiYBr$_6$:Ce was reported to have the same crystal structure, but with a smaller band gap of 5.7 eV and a higher light yield of 25,000 ph/MeV [28]. Similar Li-containing elpasolite scintillators including Cs$_2$LiLaCl$_6$:Ce, Cs$_2$LiLaBr$_6$:Ce [26], Cs$_2$LiLuCl$_6$:Ce [29], and Cs$_2$LiCeBr$_6$ [30] were reported elsewhere. Besides the Li-containing elpasolite scintillators, non-Li containing halides elpasolites were also reported as promising scintillators, such as Cs$_2$NaCeBr$_6$ [31], Cs$_2$NaGdBr$_6$:Ce [32], and Cs$_2$NaLaI$_6$:Ce [33].

Among all the reported halide elpasolite scintillators, of which none showed light yield exceeding 50,000 ph/MeV. The highest record was 50,000 ph/MeV ($^{137}$Cs source) for Cs$_2$LiLaBr$_6$:Ce with optimized Ce concentration [34]. A recent theoretical work on the rare-earth chloride elpasolites by Du *et al.* [35] showed that: 1) the localized *d* or *f* states of the trivalent rare earth form the conduction band edge while the large distance between the trivalent cations in the double-perovskite structure further localizes these states; 2) the localized Cl 3p states make up the valence band, which is narrow with small dispersion. The narrow conduction and valence bands favor charge localization, resulting in inefficient carrier mobility to the activators such as Ce$^{3+}$. Less electronegative halogens, i.e., Br and I should enhance cross-band-gap hybridization and lead to more efficient carrier (holes in particular) transport.

However, a larger ionic radius of Br$^-$ or I$^-$ lowers the Goldschmidt tolerance factor (Eq. 1) and can potentially lead to low symmetry structure. According to Zhou and Doty's [25,36] predictions on cubic halide elpasolites with the embedded-ion method, a cubic lattice can experience a symmetry breaking or structural distortion, i.e. from cubic to tetragonal, when the Goldschmidt tolerance factor is lowered to 0.909. Here the calculated Goldschmidt tolerance factor of Cs$_2$NaLaBr$_3$I$_3$ and Cs$_2$NaYBr$_3$I$_3$ are 0.902 and 0.921, respectively.

For the elpasolite bearing low symmetry crystal structures, such anisotropic structure and the solid-solid phase transition could prevent us from obtaining high quality single crystals. As a matter of fact, we failed to obtain Cs$_2$NaLaI$_6$ and Cs$_2$NaYI$_6$ single crystals during our multiple trials for crystal growth. The as-grown poly-crystals contained low symmetry phases, and had poor scintillation performance.

In the previous work, we proposed a new approach to engineer the halide elpasolite scintillators by mixing halogen anions of iodine and bromine [37]. This method has two main advantages: 1) the iodine with less electronegativity improves the charge carrier mobility efficiency; 2) more importantly, the partial mixing can still maintain cubic or nearly cubic crystal symmetry, i.e. Cs$_2$NaYBr$_3$I$_3$ and Cs$_2$NaLaBr$_3$I$_3$ preserve the cubic and tetragonal structure, respectively. The gamma ray light yield of Cs$_2$NaLaBr$_3$I$_3$: 5% Ce (by mole) is 58,000 ph/MeV. Moreover, an excellent energy resolution of 2.9% at 662 keV is achieved in small specimen. The light yield and energy resolution are better than both endpoint elpasolites of the Br-I solid solution. Cs$_2$NaYBr$_3$I$_3$: 2% Ce has an energy resolution of 3.3% at 662 keV, and a light yield of 43,000 ph/MeV [37].

The aim of this work is to investigate the origins of the scintillation emission in two intrinsic and extrinsic (Ce-doped) mixed-anion elpasolites. The joint experimental and theoretical study indicates the potential of developing intrinsic scintillators with high light yield based on STE emission at room temperature (RT). It is the first time the comprehensive scintillation properties of intrinsic mixed elpasolite crystals are studied. A variety of experimental techniques were employed to measure their spectroscopic and scintillation response at different temperatures. Electronic structures of the non-mixed elpasolite were calculated, as well as the binding energy of electron/hole polaron and self-trapped excitons (STE). The results indicate that unique electronic structure and the large binding energies of the STE are stable at RT. Therefore, the stable STE leads to the scintillation emission of the intrinsic mixed elpasolite, which is rarely observed in other metal halides at RT.

Furthermore, the method of anion mixing can be applied to improve the scintillation properties, and broaden the use of elpasolite crystals in the radiation detection applications.

## II. EXPERIMENTAL & THEORETICAL METHODS

### A. Single Crystal Growth

Single crystals of intrinsic $Cs_2NaYBr_3I_3$ and $Cs_2NaLaBr_3I_3$, and extrinsic $Cs_2NaYBr_3I_3$: Ce and $Cs_2NaLaBr_3I_3$: Ce with various Ce concentration (by mole) were grown by the Bridgman method. All the Ce-doped samples in this work will be referred as extrinsic samples. All the starting materials were 4N pure anhydrous materials purchased from Sigma Aldrich. In order to drive out the residual oxygen and moisture, the starting materials were baked in a vertical clamshell furnace under vacuum ($10^{-6}$ torr) at 250 °C for 6-15 hours before melting. Iodine in the mixed-anion elpasolite came from CsI and NaI. Then the starting materials were melted and mixed by Multiple Alternating Direction (MAD) method [38-40]. The furnace was programmed with two zones: hot zone at top, and cold zone at bottom. The pulling rate was ~3 mm/h, and the cooling rate was 3-5 °C/h. Single crystals with 8 mm to 15 mm in diameter were successfully obtained. Fig. 1 shows a crystal boule of $Cs_2NaYBr_3I_3$: Ce during the growth and after growth, as well as the crystal structure of a cubic mixed-anion elpasolite.

Inductively coupled plasma optical emission spectroscopy (ICP-OES) (Optima 2100 by PerkinElmer ®) was utilized to detect the concentration of $Ce^{3+}$. The instrument detection limit was 1 ppm. All the intrinsic crystal samples were dissolved in DI water. The standard cerium ICP reference solution was used.

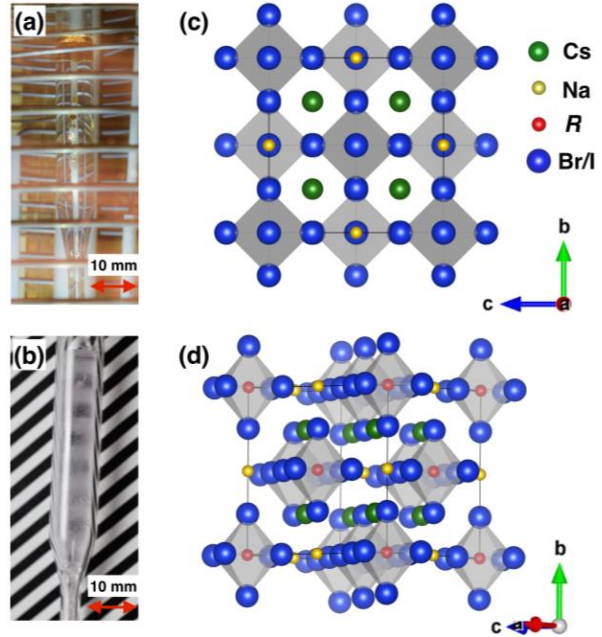

FIG. 1. (a) The crystal was grown in a gold-coated transparent furnace and (b) a transparent crystal boule of $Cs_2NaYBr_3I_3$: 2%Ce; (c-d) the cubic crystal structure of mixed-anion elpasolite viewed from two perspectives. The drawings in (c) and (d) are produced using Visualization for Electronic and Structure Analysis (Vesta) software [41].

### B. Scintillation Properties

Radioluminescence (RL) spectra were recorded at RT by exciting the samples with X-rays from a CMX-003 X-ray generator. The target material was Cu. The voltage and current of the X-ray tube were 35 kV and 0.1 mA. The emission spectra were recorded with a 150 mm focal length monochromator (PI ACTON SpectraPro SP-2155m) over a wavelength range of 200 to 800 nm. The aperture of the incident light was at maximum. The experimental geometry is the standard reflection mode, and most of the radiation interaction occurred near the surface of the crystal [42].

The scintillation decay time was measured at RT with the time correlated single photon counting technique [43]. The photomultiplier tubes (PMT) used for start and stop signals were both Hamamatsu R2059. An Ortec 556 High Voltage Power Supply (HVPS) was set to -1700 V. The measurement range was 10 µs. The irradiation source was $^{137}Cs$. The sample was next to the start PMT and 4.5 inch far away from the stop PMT. All the scintillation time curves were fit by exponential decay functions.

The scintillation light yield was measured by coupling a sample to the PMT to record the pulse height spectra. A Hamamatsu R6321-100 PMT was used for the energy resolution calculation because of its high quantum efficiency, and a calibrated Hamamatsu R3177-50 with detailed quantum efficiency at each wavelength (200-700 nm) was used for the absolute value of light yield. A $^{137}$Cs gamma ray source was used in the measurement. A Canberra 2005 pre-amp and an Ortec 672 spectroscopy amplifier with a shaping time of 10 μs amplified and shaped the PMT signal. An Ortec 556 HVPS was set to -1600V for PMT R3177-50, and -1000V for R6321-100, respectively. A multiple channel analyzer (Tukan 8K) was used to histogram the pulses [44]. A Spectralon hemispherical dome was used to reflect the scintillation light into the PMT. The sample was put into a quartz vial filled with mineral oil, which was used to protect the sample from moisture. The total light loss due to the vial was ~10%.

### C. Optical Properties

Photoluminescence (PL) emission and excitation spectra were measured with a Horiba Jobin Yvon Fluorolog 3 Spectrofluorometer equipped with a 450W Xe lamp. Horiba Jobin Yvon NanoLED light sources with various wavelengths were used for photoluminescence decay measurement; the pulse duration is less than 1 ns, and the repetition rate of the LED was set to 1 MHz. A Hamamatsu R928 PMT was used to record the emission as a function wavelength. The sample was protected in a vacuum-tight sample holder with transparent quartz window. A closed cycle compressed helium cryostat (Advanced Research Systems, DE-202) was used to cool and heat the sample from 40 K to 750 K under vacuum ($< 10^{-3}$ torr). The cooling and heating rate was set to 9K/min and controlled by a Lakeshore 332 Temperature Controller.

### D. Computational Method

Density functional calculations were performed to study electronic structure and carrier self-trapping in $Cs_2NaLaBr_6$, $Cs_2NaYBr_6$, $Cs_2NaLaI_6$ and $Cs_2NaYI_6$ [45,46]. PBE0 hybrid functionals [47], which incorporate 25% Hartree-Fock exchange, were used to calculate band structures and energetics of small polarons STE. The use of hybrid functionals provides improved description of band gaps, defects, and charge localization associated with the formation of small polarons and STE [48-51]. It is difficult to simulate a random alloy. We simply arranged the Br and I ions such that, within each $MBr_3I_3$ (M = Na or La/Y) octahedron, there is a three-fold symmetry.

The electron-ion interactions were described using projector augmented wave potentials [52,53]. The valence wave functions were expanded on a plane-wave basis with a cutoff energy of 260 eV. Experimental lattice constants were used for all elpasolites. Atomic coordinates were optimized by minimizing the Feynman-Hellmann forces to below 0.05 eV/Å.

The charge transition level $\varepsilon(q/q')$, induced by Ce impurity or polarons, is determined by the Fermi level ($\varepsilon_f$) at which the formation energies of the impurity or defect with charge states $q$ and $q'$ are equal to each other. $\varepsilon(q/q')$ can be calculated using

$$\varepsilon(\frac{q}{q'}) = \frac{E_{D,q} - E_{D,q'}}{q - q'} \quad (3)$$

Where $E_{D,q}$ ($E_{D,q'}$) is the total energy of the supercell that contains the relaxed structure of a defect at charge state $q$ ($q'$).

The binding energies of hole and electron polarons (or the energies of hole and electron polarons relative to those of free hole and free electron) are $\varepsilon_{hole-pol}(+/0)$-$\varepsilon_V$ and $\varepsilon_c$-$\varepsilon_{electron-pol}(0/-)$, respectively. Here, $\varepsilon_V$ and $\varepsilon_c$ are the energies of the valence band maximum (VBM) and the conduction band minimum (CBM), respectively.

## III. RESULTS

### A. RL Spectra

The RL emission spectra comparison of both intrinsic and extrinsic $Cs_2NaLaBr_3I_3$ and $Cs_2NaYBr_3I_3$ at RT is shown in Fig. 2. The intrinsic samples have broader emission peak compared with the extrinsic samples, which can be attributed to the STE emission. Such broad emissions were also observed in other halide elpasolites RL spectra [15,28,54]. The STE emission in the extrinsic samples are vaguely shown, this indicates a good energy transfer from the host to the $Ce^{3+}$ ions.

It is worth noting that the emission peak of the extrinsic samples looks asymmetric where there are two close peaks (the splitting in Fig. 2 (b) is more visible). These two peaks originate from the split ground states $4f$ ($^2F_{5/2}$) and $4f$ ($^2F_{7/2}$) of $Ce^{3+}$.

The quantum efficiency curve of a common Photomultiplier tube (PMT) (Hamamatsu H3177-50) is shown in Fig. 2 (c). One can see the emission of both the intrinsic and extrinsic scintillators matches well with the highly efficient detection region of the PMT. This guarantees the high photon detection efficiency during measurements.

## B. PL Spectra

The PL excitation and emission spectra of extrinsic and intrinsic $Cs_2NaLaBr_3I_3$ at 40 K are shown in Fig. 3 and Fig. 4. In the PL excitation spectra of both extrinsic and intrinsic $Cs_2NaLaBr_3I_3$, an isolated excitation band is observed from 250 nm (4.96 eV) to 280 (4.43 eV) nm in the shadowed regions of Fig. 3 (a) and (b) [55]. Compared with the extrinsic $Cs_2NaLaBr_3I_3$, this excitation band is much stronger in intrinsic $Cs_2NaLaBr_3I_3$. It is ascribed to an exciton excitation band located slightly below the conduction band for electrons [56]. The ionized electron is not free to move and could not reach the conduction band. Notice that the exciton band energy can be determined by the optical absorption/transmittance measurement, where it could be extrapolated from the fundamental absorption edge [37]. In addition, the broad excitation band from 310 nm to 405 nm is assigned to the splitting of $Ce^{3+}$ $5d$ states.

In the emission spectra of $Cs_2NaLaBr_3I_3$: Ce in Fig. 4 (a), the splitting of Ce-$4f$ levels is well resolved. The intense 420 nm and 460 nm emission peaks are attributed to the transition from $Ce^{3+}$ $5d$ state to the split ground states of $4f$ ($^2F_{5/2}$) and $4f$ ($^2F_{7/2}$). Such emissions are observed when the excitation falls in the $Ce^{3+}$ excitation band (310 to 405 nm). However, when excited with 275 nm, the emission peak becomes broader, although the $Ce^{3+}$ $4f$-splitting feature can still be seen. This emission possibly comes from a combination of STE emission and $Ce^{3+}$ emission. In the intrinsic $Cs_2NaLaBr_3I_3$ as shown in Fig. 4 (b), the $Ce^{3+}$ emission is also observed at 420 nm and 460 nm. This is probably due to the trace amount of Ce contamination in the sample, even though ICP-OES did not detect any $Ce^{3+}$ ions. When excited the nominal intrinsic $Cs_2NaLaBr_3I_3$ at 275 nm, the emission is broad, and no resolved $Ce^{3+}$ emission is observed comparing with the extrinsic elpasolites. This indicates that STE emission dominates in the intrinsic samples, while STE is suppressed with $Ce^{3+}$ doping.

The integrated PL emission intensity of both intrinsic $Cs_2NaLaBr_3I_3$ and $Cs_2NaLaBr_3I_3$: 5% Ce at different excitation wavelengths is shown in Fig. 4 (c). The integration is from 350 nm to 550 nm. For the intrinsic sample, the emission intensity is much more intense when the excitation falls in the exciton excitation band, which again indicates the STE dominates in the intrinsic samples. In the Ce-doped sample, the change of the integrated emission intensity is more stable, and the value is slight higher when the excitation falls in the $Ce^{3+}$ $4f$-$5d$ excitation band.

Similar behaviours were also observed in extrinsic and intrinsic $Cs_2NaYBr_3I_3$, as shown in Fig. 5 and Fig.

6. In the excitation spectra in Fig. 5, the long wavelength excitation bandwidth of the Ce-doped sample is larger than the intrinsic sample. The feature of exciton excitation band is observed in both samples. In the emission spectra of intrinsic $Cs_2NaYBr_3I_3$ in Fig. 6(a), when excited with shorter wavelength of 273 nm, a distinct broader long wavelength emission peak occurs, comparing with the well-defined Ce-doped emission in Fig. 6 (b).

Fig. 7 shows the intrinsic $Cs_2NaYBr_3I_3$ PL spectra at RT for comparison. Different from the 40 K excitation spectra, the short wavelength exciton excitation band (250- 280 nm) is enhanced with elevated temperature compared with $Ce^{3+}$ excitation band (330- 410 nm). Based on the emission spectra in Fig. 7 (b), the STE-induced emission ranges broadly from 320 nm to 550 nm, and it is similar to its RL emission spectrum. On the other hand, the $Ce^{3+}$ excitation band overlaps with the broad STE emission band, and this can result in radiative transfer from STE to $Ce^{3+}$, i.e., the STE emission can be absorbed at the $Ce^{3+}$ site which re-emits photons [28].

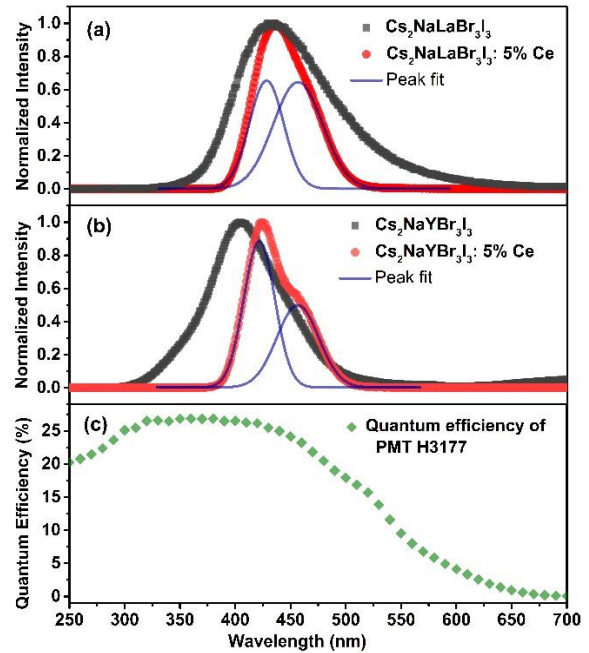

FIG. 2. RL spectra comparison of intrinsic and extrinsic samples of (a) $Cs_2NaLaBr_3I_3$ and (b) $Cs_2NaYBr_3I_3$. The emission peaks of the extrinsic samples are fit with Gaussian function, as shown in the blue curves. The data is normalized by the maximum peak intensity. (c) The quantum efficiency curve of a common PMT (Hamamatsu H3177-50). The spectra of extrinsic samples $Cs_2NaLaBr_3I_3$ and $Cs_2NaYBr_3I_3$ are adapted from [37].

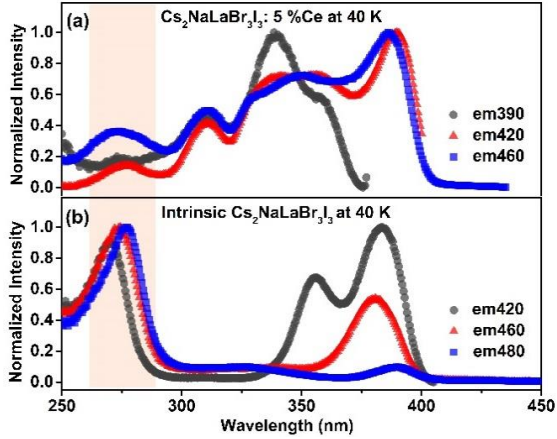

FIG. 3. PL excitation spectra at 40 K of (a) Cs$_2$NaLaBr$_3$I$_3$: 5% Ce and (b) intrinsic Cs$_2$NaLaBr$_3$I$_3$. Both spectra are normalized to the maximum peak. Excitation spectra are monitored at various emission wavelengths. The highlighted region from 250 nm to 280 nm indicates the strong exciton excitation band in intrinsic samples.

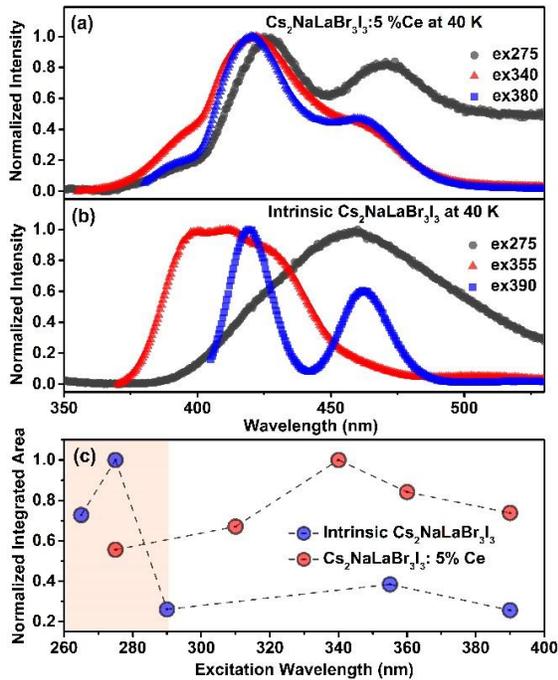

FIG. 4. PL emission spectra at 40 K of (a) Cs$_2$NaLaBr$_3$I$_3$: 5% Ce and (b) intrinsic Cs$_2$NaLaBr$_3$I$_3$. Both spectra are normalized to the maximum peak. (c) The integrated emission spectra under different excitation wavelengths of intrinsic Cs$_2$NaLaBr$_3$I$_3$ (blue dot) and extrinsic Cs$_2$NaLaBr$_3$I$_3$: 5% Ce (red dot), respectively. The integration range is from 350 nm to 550 nm. The integrated intensity was calculated with the un-normalized raw data. The raw intensity comparison without normalization can be found in the Supplemental Materials [55]. The intrinsic sample gives strong emission intensity under the exciton excitation at shorter wavelength.

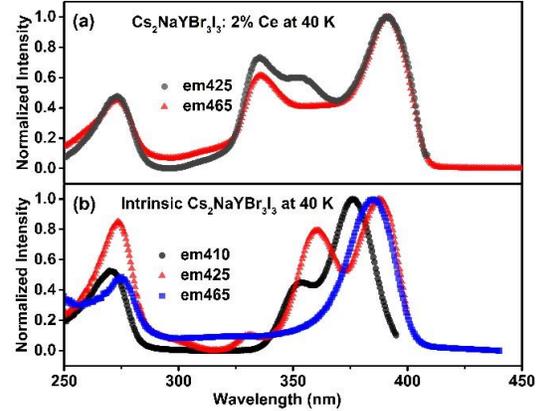

FIG. 5. PL excitation spectra at 40 K of (a) Cs$_2$NaYBr$_3$I$_3$: 5% Ce, and (b) intrinsic Cs$_2$NaYBr$_3$I$_3$ (b). Both spectra are normalized to the maximum peak.

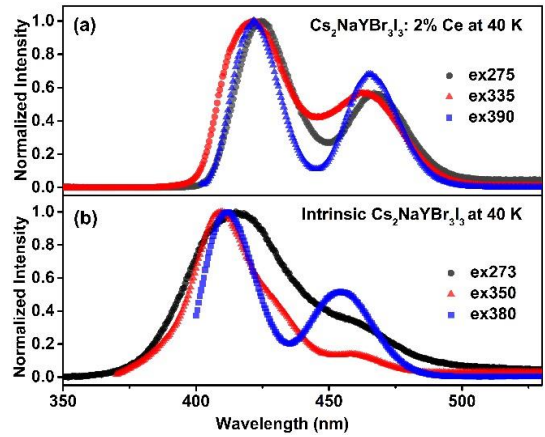

FIG.6. PL emission spectra at 40 K of (a) Cs$_2$NaYBr$_3$I$_3$: 5% Ce and (b) intrinsic Cs$_2$NaYBr$_3$I$_3$. Both spectra are normalized to the maximum peak.

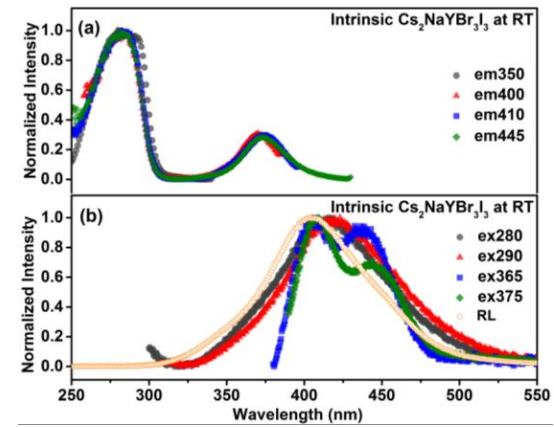

FIG. 7. (a) PL excitation and (b) PL emission spectra of intrinsic Cs$_2$NaYBr$_3$I$_3$ at RT, the RL spectrum of Cs$_2$NaYBr$_3$I$_3$ at RT is also plotted for comparison.

## C. PL Kinetics

The PL decay time of both extrinsic and intrinsic Cs₂NaLaBr₃I₃ and Cs₂NaYBr₃I₃ were recorded at different temperatures. For extrinsic Cs₂NaLaBr₃I₃ and Cs₂NaYBr₃I₃, the exciton excitation wavelengths of 295 nm, and the Ce³⁺ excitation wavelength of 370 nm were chosen respectively, in order to monitor the emissions from STE and Ce³⁺. From the PL spectra in Fig. 3 to Fig. 6, one can see the STE and Ce³⁺ emissions have broad overlap between 320 nm and 550 nm. To reveal the excitation spectra of both STE and Ce³⁺, the emission at 420 nm was monitored. Unfortunately, for intrinsic Cs₂NaLaBr₃I₃ and Cs₂NaYBr₃I₃, when using an airtight sample holder on the cryogenic station for low temperature measurement, the emission is too weak to observe when excited at 295 nm. Instead, the RT PL decay curves are shown here to illustrate the kinetics of exciton-excitation induced emission. The instrumental response decay curve is also plotted for reference. All the PL decay-fitting parameters are shown in Table 1.

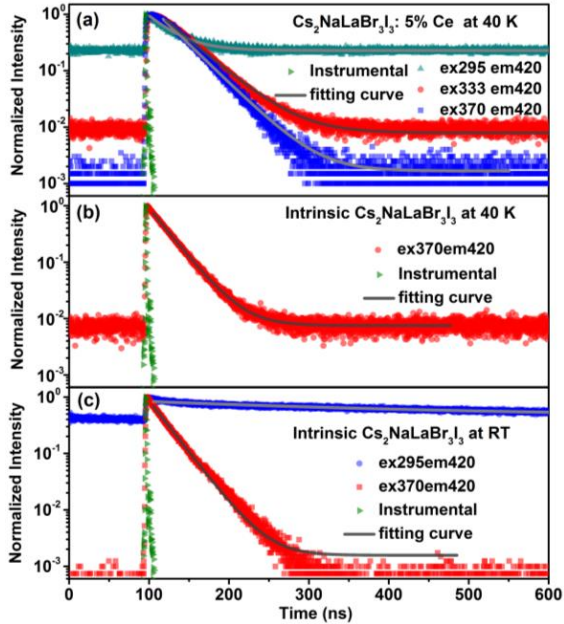

FIG. 8. (a) PL decay of Cs₂NaLaBr₃I₃: 5% Ce at 40 K: the emission at 420 nm was monitored with excitation wavelengths of 295 nm, 333 nm, and 370 nm. The instrumental response is measured to be less than 1 ns, which can be ignored in the decay fitting. The PL decay curves are fit with single decay exponential function. PL decay of intrinsic Cs₂NaLaBr₃I₃ at (b) 40 K and (c) RT. The emission at 420 nm was monitored with excitation wavelengths of 295 nm and 370 nm. The emission cannot be detected with 295 nm excitation in the 40 K measurement because the cryogenic sample holder blocked the weak emission.

In Fig. 8 (a) and Fig. 9 (a), for both extrinsic Cs₂NaLaBr₃I₃ and Cs₂NaYBr₃I₃ at 40 K, when exciting with 333 nm and 370 nm, the PL decay time is around 30 ns. Both of the excitation wavelengths belong to the Ce³⁺ excitation band, therefore, the PL decay can be ascribed to the Ce³⁺ characteristic 5d-4f transition. When exciting with 295 nm, which falls into the exciton excitation band, the PL decay monitored at 420 nm is also around 30 ns. It is clearly seen that the 295 nm-excited emission creates much higher background than the 370 nm-excited fast emission. This reveals the existence of a much longer decay component [57]. Considering the excitation wavelength of 295 nm falls in the exciton excitation band, it is reasonable that the excitation at 295 nm creates the excitons that have long lifetime in microsecond level, which is prominently longer than direct electron-hole capture of Ce³⁺.

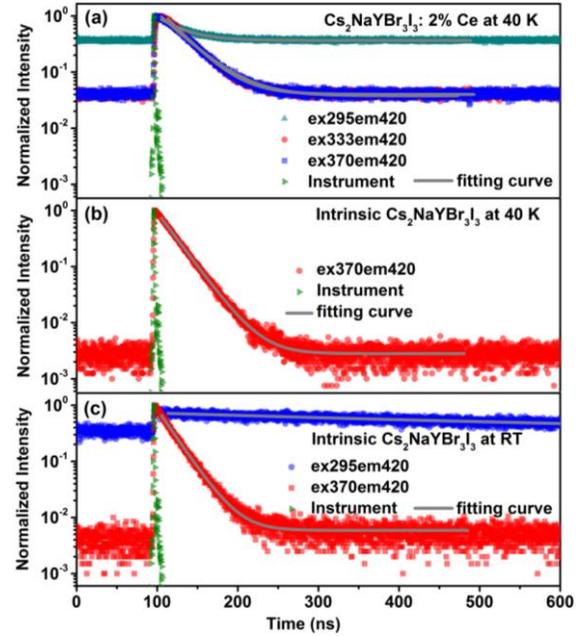

FIG. 9. (a) PL decay of Cs₂NaYBr₃I₃: 2% Ce at 40 K: the emission at 420 nm was monitored with excitation wavelengths of 295 nm, 333 nm, and 370 nm. The PL decay curves are fit with single decay exponential function. PL decay of intrinsic Cs₂NaYBr₃I₃ at (b) 40 K and (c) RT. The emission at 420 nm was monitored with excitation wavelengths of 295 nm and 370 nm. The emission cannot be detected with 295 nm excitation in the 40 K measurement because the cryogenic sample holder blocked the weak emission.

The PL decay of intrinsic samples is shown in Fig. 8 (b-c) and Fig. 9 (b-c). When exciting with 370 nm and monitoring at 420 nm, both Cs₂NaLaBr₃I₃ and Cs₂NaYBr₃I₃ show a 30 ns characteristic fast Ce³⁺ decay, similar to the extrinsic samples. This is due to the trace amount of $Ce^{3+}$ in the nominal intrinsic samples. When monitoring the 420 nm emission with 295 nm excitation, which belongs to the exciton- excitation band, the long decay time above 1 µs is observed. It is readily ascribed to the characteristic STE decay [23, 26-29, 54].

TABLE. 1. The PL decay fitting parameters of intrinsic and extrinsic $Cs_2NaLaBr_3I_3$/ $Cs_2NaYBr_3I_3$ at various emission/excitation wavelengths and temperatures.

| | PL decay parameters (ns) | | |
|---|---|---|---|
| | *ex295 em 420* | *ex333 em420* | *ex370 em420* |
| $Cs_2NaLaBr_3I_3$: 5% Ce at 40 K | 29.76±0.21 | 38.01±0.07 | 37.42±0.10 |
| Intrinsic $Cs_2NaLaBr_3I_3$ at 40 K | | | 24.83±0.03 |
| Intrinsic $Cs_2NaLaBr_3I_3$ at RT | 32.19±1.67 (16%) 1055.87±71.89 (84%) | | 23.91±0.03 |
| | | | |
| $Cs_2NaYBr_3I_3$: 2% Ce at 40 K | 30.93±0.17 | 29.35±0.05 | 32.55±0.07 |
| Intrinsic $Cs_2NaYBr_3I_3$ at 40 K | | | 22.49±0.02 |
| Intrinsic $Cs_2NaYBr_3I_3$ at RT | 31.89±0.24 (5%) 813.78±120 (95%) | | 20.47±0.03 |

For the mixed elpasolites, the $Ce^{3+}$ has a fast PL decay time of around 30 ns, while the STE has a longer decay time of more than 1 µs. Because of the domination of $Ce^{3+}$ in the emission, only in the intrinsic sample, the STE decay can be recorded distinctively. This indicates the competition between STE and $Ce^{3+}$ in the scintillation process: 1) The STE transfers its energy to $Ce^{3+}$ radiatively, which means $Ce^{3+}$ is absorbing the emission from STE. In this case, the decay time of $Ce^{3+}$ emission should be equivalent or slightly slower than the decay time of STE. 2) The STE transfers its energy to $Ce^{3+}$ non-radiatively by thermal activated diffusion. In this case, the time constant should be relatively close to the characteristic decay time of $Ce^{3+}$. One would also expect an increase of $Ce^{3+}$ emission intensity as temperature increases before reaching thermal quenching.

### D. Scintillation Kinetics

The scintillation decay between intrinsic and extrinsic samples at RT is compared in Fig. 10 (a) and (b). The decay curves of extrinsic and intrinsic $Cs_2NaLaBr_3I_3$ and $Cs_2NaYBr_3I_3$ were fit with three and two exponential decay functions, respectively. The decay time and the ratio are shown in the inset tables.

The fast decay component below 100 ns is a characteristic of $Ce^{3+}$ de-excitation process, and it is the major contribution in both extrinsic $Cs_2NaLaBr_3I_3$ and

$Cs_2NaYBr_3I_3$. It is clearly seen that the fast decay component is absent in the intrinsic samples. Instead, the slow decay component around 1 µs contributes to more than 80% of the total emission in the intrinsic samples. In both intrinsic and extrinsic samples, this microsecond slow decay component is observed and ascribed to STE. Combes *et al.* [15] and van't Spijker *et al.* [58] suggested that in the halide elpasolites, the creation of free electrons in the conduction band and free holes in the valence band is followed by the creation of self-trapped holes, i.e. $V_k$ center. The formation of $V_k$ centers is common in halides due to the localized valence band states and their soft lattice [59,60]. The self-trapped holes can trap free electrons to form self-trapped excitons, which will result in STE emission. The STE is thermally activated and can be quenched with elevated temperature. In most elpasolites, the STE can be formed at RT [15,29-31,54,61].

The intermediate decay component of several hundred nanoseconds is observed in both intrinsic and extrinsic samples. However, this decay component is not observed in the PL measurement. It is not likely to be the direct de-excitation of $Ce^{3+}$. Compared to many other $Ce^{3+}$ doped elpasolite, this intermediate decay component is commonly seen [15,29-31,54,61], yet no clear origin has been given at this point. One possible origin is the non-radiative energy transfer from STE to $Ce^{3+}$, and it is normally faster than the radiative transfer. Another possible reason can be related to the de-trapped electrons from shallow traps.

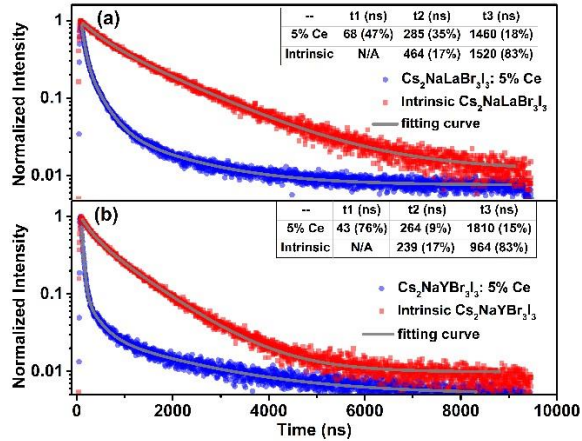

FIG. 10. Scintillation decay profiles of intrinsic and extrinsic samples of (a) $Cs_2NaLaBr_3I_3$ and (b) $Cs_2NaYBr_3I_3$. The extrinsic decay curves are fit with three decay exponential function, and the intrinsic decay curves are fit with two decay exponential function. The fitting curves are shown in solid grey lines. The scintillation rising time of the extrinsic $Cs_2NaLaBr_3I_3$ and $Cs_2NaYBr_3I_3$ can be found at [55].

### E. Scintillation Light Yield

The pulse height spectra of both intrinsic and extrinsic $Cs_2NaLaBr_3I_3$ and $Cs_2NaYBr_3I_3$ are shown in Fig. 11 (a) and (b). For the extrinsic scintillators, the Ce concentration was optimized for best energy resolution at 662 keV in our previous work [37]. Table 2 is a list of the light yield and energy resolution for selected samples.

TABLE 2. Light yield and energy resolution comparison

| | Light yield (ph/MeV) | Energy resolution (662 keV) |
|---|---|---|
| Intrinsic $Cs_2NaLaBr_3I_3$ | 39,000 | 6.6% |
| $Cs_2NaLaBr_3I_3$: 5%Ce | 58,000 | 2.9% |
| Intrinsic $Cs_2NaYBr_3I_3$ | 40,000 | 4.3% |
| $Cs_2NaYBr_3I_3$: 2%Ce | 43,000 | 3.3% |

While the intrinsic samples have lower light yield than the extrinsic samples, the intrinsic mixed elpasolites have light yield comparable to NaI:Tl. In fact, their light yield is higher than many other well-known extrinsic scintillators, such as LSO:Ce [6], YAP:Ce [62] etc.

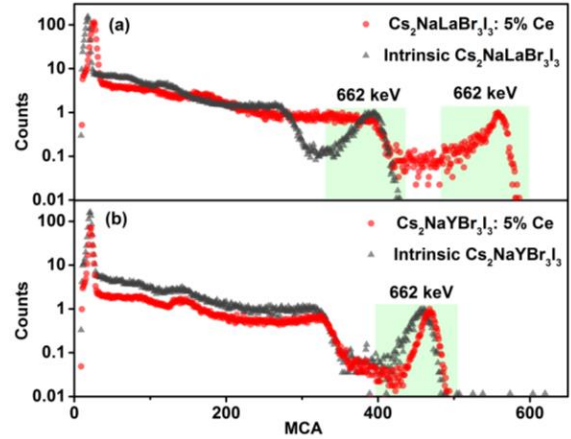

FIG. 11. Gamma ray pulse height spectra of intrinsic and extrinsic samples of (a) $Cs_2NaLaBr_3I_3$ and (b) $Cs_2NaYBr_3I_3$. The photopeak at 662 keV is highlighted to better illustrate the position. Cs-137 source was used. The pulse height spectra of extrinsic samples $Cs_2NaLaBr_3I_3$ and $Cs_2NaYBr_3I_3$ are adapted from [37].

### F. First-principle Calculation

The valence band of the rare-earth elpasolites studied here is made up of halogen p states while the conduction band is derived from the rare-earth d states. The band structures of $Cs_2NaLaBr_6$ and $Cs_2NaYBr_6$ are shown in Fig. 12. Both valence and conduction bands are narrow, having small dispersion. Narrow valence bands are typical for halides.

However, the narrow conduction band is unusual and is related to the structure and chemistry of elpasolites [35,63,64]. In rare-earth elpasolites, such as $Cs_2NaYBr_6$, the rare-earth cation is much more electronegative than the alkali metal cations and, as a result, the conduction band is mainly a rare-earth d band, which is separated in energy from the alkali metal s band. The large nearest-neighbor distance between the rare-earth cations leads to weak coupling between the rare-earth d orbitals and consequently a very narrow conduction band as seen in Fig. 12 (a) and (b).

The band gaps of $Cs_2NaLaBr_6$ and $Cs_2NaYBr_6$ calculated using PBE0 hybrid functionals are 6.31 eV and 6.25 eV. Mixing bromides with iodides in 1:1 ratio reduces the band gaps to 5.41 eV and 5.15 eV for $Cs_2NaLaBr_3I_3$ and $Cs_2NaYBr_3I_3$, in agreement with experimentally measured band gaps of 4.92 eV and 4.87 eV, respectively. Pure iodides (i.e. $Cs_2NaLaI_6$ and $Cs_2NaYI_6$) have not been synthesized. We optimized the lattice constants of $Cs_2NaLaI_6$ and $Cs_2NaYI_6$ in cubic structures and calculated the band gaps. Note that the structures of iodides are likely not cubic. The purpose of the calculations is to have a rough idea of the band gaps of iodides. The calculated band gaps of the hypothetical cubic Cs$_2$NaLaI$_6$ and Cs$_2$NaYI$_6$ are 5.02 eV and 4.93 eV, respectively. It therefore appears that mixing bromides and iodides in 1:1 ratio reduces the band gaps of the alloys substantially from those of bromides to very close to those of iodides. The substantial reduction of the band gap by alloying leads to significant increase in light yield as observed experimentally.

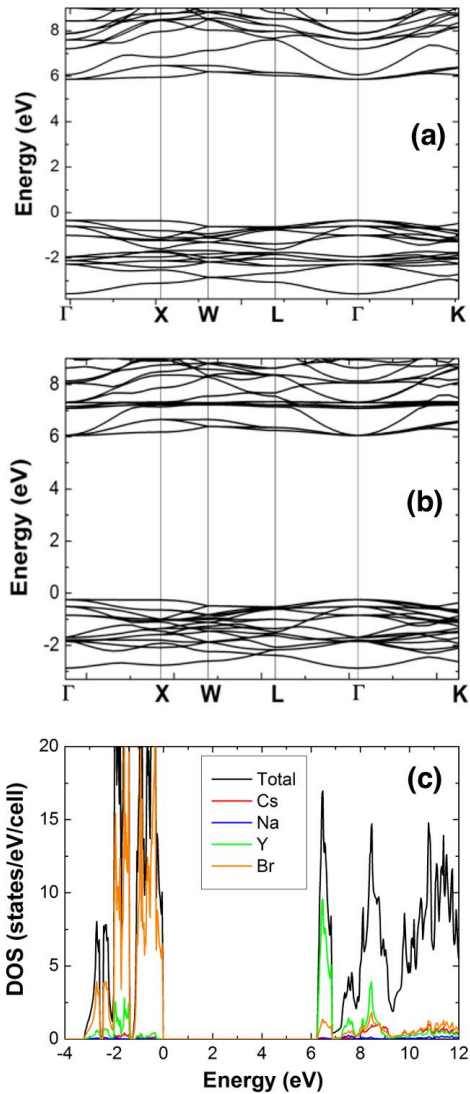

FIG. 12. Band structure of (a) Cs$_2$NaLaBr$_6$ and (b) Cs$_2$NaYBr$_6$; (c) Density of States (DOS) of Cs$_2$NaYBr$_6$.

The narrow valence and conduction bands favor the self-trapping of both holes and electrons, forming small hole and electron polarons. The calculated binding energies of small hole and electron polarons and STEs in Cs$_2$NaLaBr$_6$ and Cs$_2$NaYBr$_6$ are shown in Table 3. Note that the binding energy of a STE is calculated relative to the energies of small hole and electron polarons. The large binding energies shown in Table 3 show that STEs are stable at RT and could survive at even higher temperatures. As a result, STE emission should be observed at RT. The energy transfer in these elpasolites is due to hopping of localized STEs rather than the diffusion of free carriers. Therefore, the electron transfer is inefficient, which leads to relatively slow scintillation decay as also observed experimentally.

Mixing bromides with iodides is expected to reduce the small hole binding energy and the STE binding energy, resulting in somewhat faster energy transfer.

TABLE 3. Calculated binding energies (in eV) of small hole and electron polarons and STE in Cs$_2$NaLaBr$_6$ and Cs$_2$NaYBr$_6$

|  | Hole polaron | Electron polaron | STE |
|---|---|---|---|
| Cs$_2$NaLaBr$_6$ | 0.63 | 0.47 | 0.42 |
| Cs$_2$NaYBr$_6$ | 0.51 | 0.39 | 0.36 |

## IV. DISCUSSION

An energy diagram of Cs$_2$NaLaBr$_3$I$_3$: Ce and Cs$_2$NaYBr$_3$I$_3$: Ce is plotted in Fig. 13 based on the well-resolved PL excitation spectra and temperature dependent PL decay time of Ce$^{3+}$ [55]. From the temperature dependent photoluminescence kinetics measurement, it is found that the estimated $5d_1$ level is located more than 1 eV below the conduction band in both compounds. This region could potentially be occupied by electron traps. The trapped electrons can then be thermally de-trapped with time constants related to the trap depths. Afterwards, the de-trapped electron can recombine with a hole, which is previously trapped at Ce$^{3+}$ site. This delayed process may cause the intermediate scintillation decay (a few hundred nanosecond). Thermoluminescence studies are necessary to determine the lifetimes of the electron traps, and potentially correlate them to the intermediate scintillation decay time.

Based on the PL decay and scintillation decay results, three scintillation mechanisms are proposed in the mixed-anion elpasolites, as shown in Fig. 14.

1) Fast emission: After initial ionization of free holes and electrons, the Ce$^{3+}$ luminescence centers sequentially capture holes from valence band and then electrons from conduction band and, then de-excite via photon emission.

2) Intermediate emission: Shallow defects temporarily trap electrons during the electron thermalization stage. The trapped electrons can be de-trapped thermally and then migrate to $Ce^{3+}$ centers.

3) Slow emission: After a hole is created, it can be trapped by two anions in the valence band and form a $V_k$ center. The $V_k$ center can trap a free electron to form a STE. The STE is capable of radiative de-excitation and results in photon emission. The emitted photon can either escape from the crystal surface or be absorbed by $Ce^{3+}$ with subsequent re-emission. Alternatively, the STE can migrate to a $Ce^{3+}$ site and transfer its energy to the $Ce^{3+}$ non-radiatively.

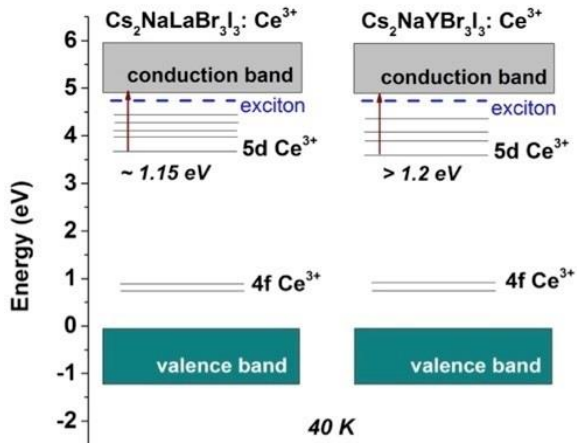

FIG. 13. Energy diagram of $Cs_2NaLaBr_3I_3$ (left) and $Cs_2NaYBr_3I_3$ (right) at 40 K.

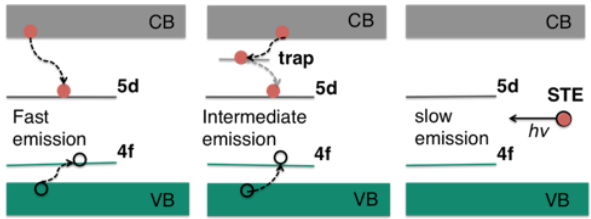

FIG. 14. Diagram of three different scintillation decay processes in the mixed elpasolite scintillators.

$Cs_2NaLaBr_3I_3$ and $Cs_2NaYBr_3I_3$, are discussed in this work. Furthermore, the established connection between the theoretical and experimental results on these two compounds can be developed as a predictive model on scintillator design. For instance, in the band engineering of elpasolite halides, the correlation between mixing anion and scintillation properties can be predicted: less electronegative halogen is preferred until the distortion of lattice severely deteriorate the crystal quality (low transparency, cracks, secondary phase etc.). Also, the cations can also play an important role. They can change the crystal structure (e.g. $Cs_2NaLaBr_3I_3$ is cubic and $Cs_2NaYBr_3I_3$ is tetragonal), alter the conduction band edge (especially the tri-valence cations), and eventually affect the scintillation properties.

## V. CONCLUSIONS

The spectroscopic analyses and scintillation properties of intrinsic and extrinsic $Cs_2NaYBr_3I_3$ and $Cs_2NaLaBr_3I_3$ mixed-anion elpasolites were discussed. Compared with the intrinsic scintillators, the energy resolution (at 662 keV) of the extrinsic scintillators is improved by 56% and 23% for $Cs_2NaLaBr_3I_3$ and $Cs_2NaYBr_3I_3$, respectively. The PL excitation and emission spectra indicate that the exciton excitation band is below the optical absorption edge, which results in a broad STE emission overlapping with the $Ce^{3+}$ emission. The PL decay time of the STE emission is about 1 μs compared with the 30 ns decay time of $Ce^{3+}$ 5d-4f transition at 40 K. Ab initio calculations performed on $Cs_2NaLaBr_6$ and $Cs_2NaYBr_6$ show their small dispersive conduction bands, which can lead to stable electron polarons at RT. The large binding energy of STE suggests its stability at RT, thus results in the scintillation emission directly from STE. The calculated large STE binding energies are consistent with the observed high light yield of the intrinsic samples due to STE emission. Mixing less electronegative iodine with bromine can effectively reduce the STE binding energy and band gap of the host materials. This could improve the energy transfer efficiency from STE to $Ce^{3+}$ in the extrinsic scintillators. Mixing anions of halide scintillators can be an effective approach to improve the performance of current in-use scintillators. Also, it can be used to design new scintillators to meet the specific need in radiation detection applications.

## ACKNOWLEDGEMENTS

We thank Mr. Bo Bishop (University of Tennessee-Knoxville) for the technical support of making quartz ampoules, and Dr. Pieter Dorenbos (Delft University of Technology) for the scientific discussions. Mao-Hua Du and Hongliang Shi are supported by the Department of Energy, Office of Science, Basic Energy Sciences, Materials Sciences and Engineering Division.